\documentclass[aps,prb,10pt,twocolumn,superscriptaddress,floatfix]{revtex4-1}

\usepackage{graphicx}
\usepackage[dvipsnames]{xcolor}
\colorlet{BLUEVIOLET}{BlueViolet}
\usepackage{amsmath}
\usepackage{ifsym}
\usepackage{amssymb}
\usepackage{colortbl}
\usepackage{epstopdf}
\usepackage{lettrine}
\usepackage{type1cm}

\makeatletter

\newcommand{\Rmnum}[1]{\expandafter\@slowromancap\romannumeral #1@}
\makeatother

\usepackage[bf, sf, justification=raggedright]{caption}
\DeclareCaptionLabelSeparator{line}{ $|$ }
\usepackage{xr}
\begin{document}

\title{\textsf{Frustration and thermalisation in an artificial magnetic quasicrystal}}

\author{Dong~Shi}
\affiliation{School of Physics \&\ Astronomy, University of Leeds, Leeds LS2 9JT, United Kingdom}

\author{Zoe~Budrikis}
\affiliation{Institute for Scientific Interchange Foundation, Via Alassio 11/C, 10126 Torino, Italy}

\author{Aaron~Stein}
\affiliation{Center for Functional Nanomaterials, Brookhaven National Laboratory, Upton NY 11793, USA}

\author{Sophie~A.~Morley}
\affiliation{School of Physics \&\ Astronomy, University of Leeds, Leeds LS2 9JT, United Kingdom}

\author{Peter~D.~Olmsted}
\affiliation{School of Physics \&\ Astronomy, University of Leeds, Leeds LS2 9JT, United Kingdom}
\affiliation{Department of Physics, Institute for Soft Matter Synthesis and Metrology,
Georgetown University, 37th and O Streets, N.W., Washington, D.C. 20057, USA}

\author{Gavin~Burnell}
\affiliation{School of Physics \&\ Astronomy, University of Leeds, Leeds LS2 9JT, United Kingdom}

\author{Christopher~H.~Marrows}\email[email:~]{c.h.marrows@leeds.ac.uk}
\affiliation{School of Physics \&\ Astronomy, University of Leeds, Leeds LS2 9JT, United Kingdom}


\begin{abstract}
\textbf{Artificial frustrated systems offer a playground to study the emergent properties of interacting systems. Most work to date has been on spatially periodic systems, known as artificial spin ices when the interacting elements are magnetic. Here we have studied artificial magnetic quasicrystals based on quasiperiodic Penrose tiling patterns of interacting nanomagnets. We construct a low energy configuration from a step-by-step approach that we propose as a ground state. Topologically induced emergent frustration means that this configuration cannot be constructed from vertices in their ground states. It has two parts, a quasi-one-dimensional ``skeleton'' that spans the entire pattern and is capable of long-range order, surrounding ``flippable'' clusters of macrospins that lead to macroscopic degeneracy. Magnetic force microscopy imaging of Penrose tiling arrays revealed superdomains that are larger for more strongly coupled arrays, especially after annealing the array above its blocking temperature.} 
\end{abstract}

\date{\today}
\maketitle

\lettrine[lines=3,lraise=0.2,loversize=-0.2]{\textcolor{gray}{\sf G}}{}eometrical frustration not only exists in crystalline materials such as the Ice I$_\mathrm{h}$ phase of water and the rare-earth pyrochlore spin ices \cite{1933Bernal,1997Harris}, but can also be realised in a wide range of artificial systems \cite{2006Wang,2008Han,2013Latimer,2016Ortiz}. Since they are built using nanotechnology, the structure of such a system, and the frustrated interactions it embodies, can be designed rather than discovered. When magnetic, such systems are known as artificial spin ices \cite{2013Nisoli}.

\begin{figure}[b]
  \includegraphics[width=8cm]{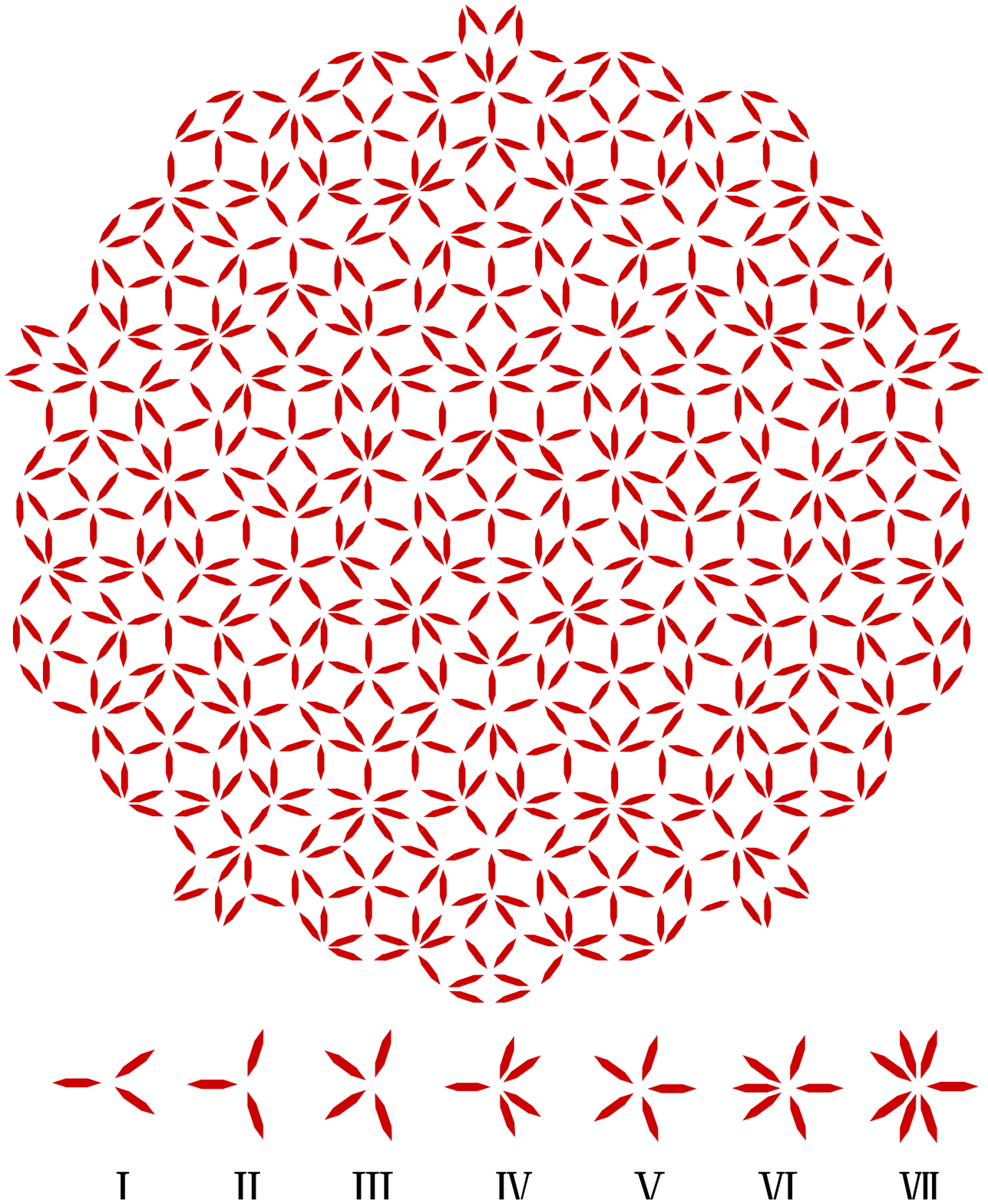}
  \caption{\textsf{\textbf{Penrose tiling pattern and the seven types of vertex found within it.} Our pattern is of the P3 type, the central portion is shown here. Each vertex type within it has a distinct geometry and co-ordination number. Note: the angles between the three islands in type II vertex are not all equal.}}\label{fig:fig1}
\end{figure}

Two different analogs of the pyrochlores are the square \cite{2006Wang,2011Morgan,2013Morgana,2013Porro,Farhan2013b,2014Kapaklis,2016perrin} and kagom\'{e} \cite{2005Tanaka,2008Qi,2010Ladak,2011Mengotti,2011Rougemaille} ice arrays, which have been widely studied. The square ice array has a long-range ordered antiferromagnetic ground state that is easily accessible through annealing \cite{2011Morgan,2013Porro,2013Zhang}, although introducing a height offset between islands can restore the extensive degeneracy \cite{2016perrin}. Meanwhile the kagom\'{e} pattern has stronger frustration and a richer phase diagram \cite{2009Moller,2011Chern}, whose phase transitions have been probed by low energy muon spectroscopy \cite{2015Anghinolfi}. Recent attention has focussed on thermal excitations in these systems \cite{2007Nisoli,2011Morgan,2011Budrikis,2012Kapaklis,2013Porro,2013Zhang,Farhan2013a,Farhan2013b,Chioar2014,2014Kapaklis,2015Drisko,2017Morley}, as well as new lattices designed to give rise to novel phenomena \cite{2013Morrison}. These include the `shakti' lattice, which displays topologically induced emergent frustration \cite{2014Gilbert}, the `tetris' lattice, which exhibits an emergent reduction in dimensionality \cite{2016Gilbert}, and artificial charge ices suitable for data storage \cite{2016Wang}.

These arrays are spatially periodic with discrete translational symmetry. Since artificial systems may be designed arbitrarily, there is no need to always respect this constraint. Indeed, the discovery of quasicrystalline materials \cite{1984Schectman} shows that nature does not always respect it either. Quasicrystals have structures that lack translational symmetry \cite{1992Janot}, and so possess complex forms of magnetic frustration leading to glassy behaviour \cite{1998Charrier,1998Islam,2005Sato}. Building artificial magnetic quasicrystals based on Penrose tilings brings the ability to inspect the magnetic configuration at the level of individual macrospins, allowing deeper insight into this frustration and its manifestations. Kite-and-dart tilings (which have tile edges of different lengths) with connected magnetic elements have been studied\cite{2013Bhat,2016Farmer,Brajuskovic_2016}.

Here we have fabricated and studied a Penrose tiling array based on the thick and thin rhombus tiles with discrete islands. The central portion of the pattern is depicted in Fig.~\ref{fig:fig1}, along with the seven vertex types (denoted Type~\Rmnum{1} to Type~\Rmnum{7}) from which it may be constructed. The rhombus tiles have a single edge length, so all our islands have the same nominal volume and hence the same paramagnetic blocking temperature, important for defining a simple thermal anneal protocol. The fact that out islands are not connected means that we are able to tune the coupling strength by varying their spacing. We have determined a set of degenerate lowest energy configurations based nearest neighbour interactions, in which the macrospins become separated into two groups. A `skeleton' spans the lattice, forming a framework with a unique (up to a global spin-flip) ground state. This surrounds `flippable' groups of macrospins that give rise macroscopic degeneracy. These features resemble those arising from Monte Carlo simulations (and magnetomechanical mesoscopic realisations) of XY spins on a quasiperiodic lattice \cite{2003Vedmedenko}. Topologically induced emergent frustration is present \cite{2014Gilbert}. We compared three different protocols for thermalising this system: one-shot annealing during fabrication to produce a thermalised as-grown state \cite{2011Morgan,2012Nisoli}, AC-demagnetisation \cite{2006Wang,Wang2007}, and thermal annealing \cite{2013Porro,2013Zhang}. Effective thermalisation leads to superdomains within the skeleton that are separated by Ising-like walls. The superdomains are largest for an annealed sample with strong interisland coupling. 

\section*{Ground state of the artificial magnetic quasicrystal}

\begin{figure*}[t]
  \includegraphics[width=12cm]{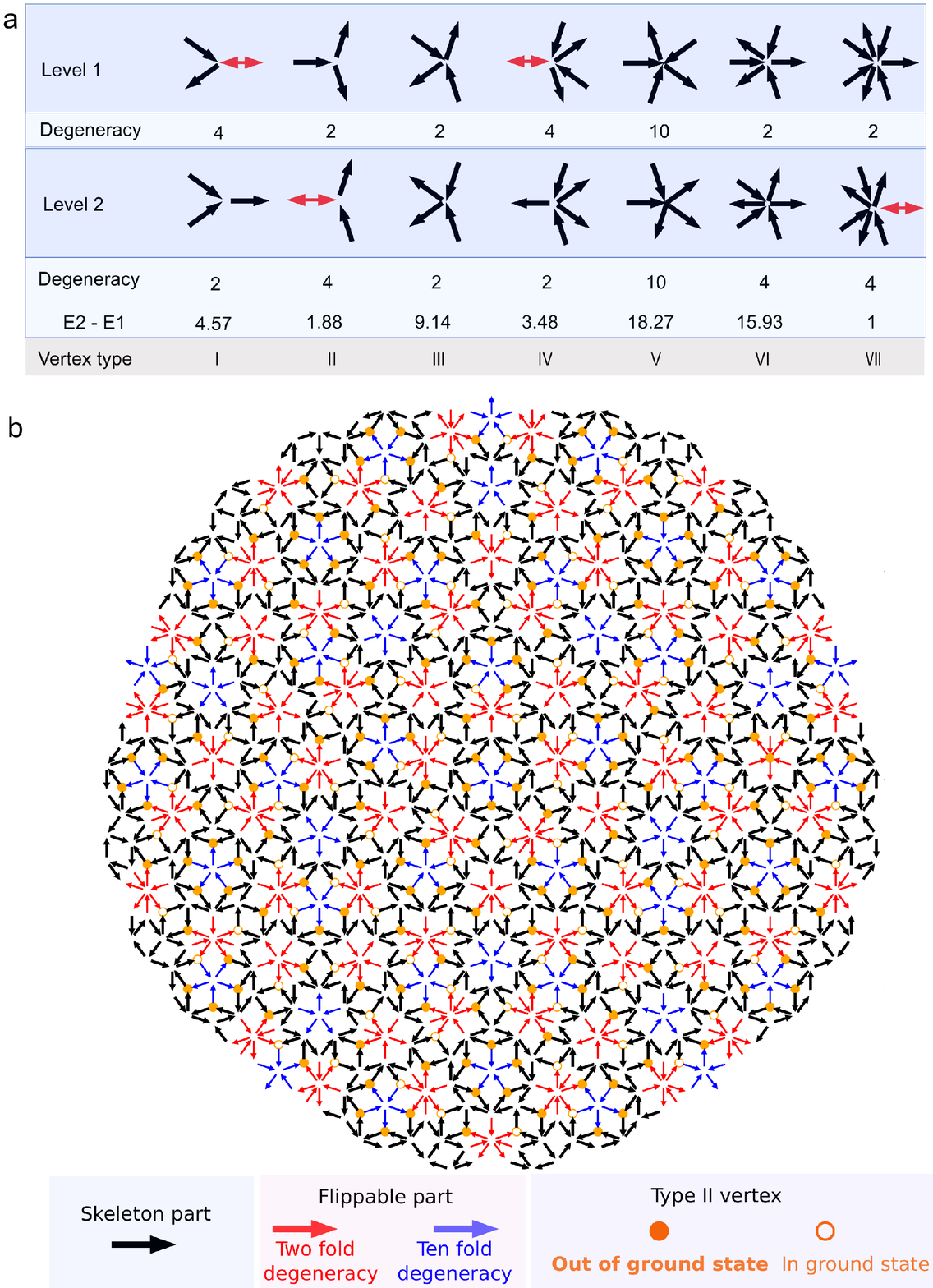}
  \caption{\textsf{\textbf{The lowest energy configuration of each vertex and a low-energy state of the entire pattern.} \textbf{a}, the lowest (level 1) and second lowest (level 2) energy configuration for each type of vertex in isolation are shown. Arrows drawn for each vertex represent the direction of that macrospin. The red double-headed arrows denote the moments can point to either of the two directions without changing the total vertex energy. \textbf{b}, a low energy configuration of the entire Penrose tiling pattern, constructed by joining the vertices into decagons and then combining the decagons. The black arrows represent the magnetically rigid ``skeleton'', which has a unique ground state (up to time reversal). The red and blue arrows represent ``flippable'' clusters of macrospins that have degenerate configurations within the array: twofold degenerate for red, tenfold for blue. The orange circles represent the Type~\Rmnum{2} vertices that can be forced out of their vertex-level ground state by topologically-induced emergent frustration. Those that are filled indicate a possible configuration of excited vertices, the open ones are those that have remained in their vertex-level ground state in this particular case.}}\label{fig:fig2}
\end{figure*}

We fabricated artificial magnetic quasicrystals using the P3 Penrose tiling pattern, of which Fig.~\ref{fig:fig1} shows the central portion. The pattern was generated by a `cut and project' method, projecting a five-dimensional hypercubic lattice into the two-dimensional space of our substrate plane, described in more detail in the Supplementary Information. The pattern was realised using electron beam lithography followed by electron beam evaporation of Permalloy (Ni$_{80}$Fe$_{20}$). The island size was  fixed at 450~nm long, 80~nm wide and 25~nm thick, but the intervertex distance $a$ ranged from 600 to 1000~nm to vary the coupling strength. Each island is shuttle-shaped, tipped with sharp points to allow them to be brought close together in high co-ordination vertices: further details are in the Supplementary Information. Magnetic force microscope (MFM) images demonstrated that all the islands were single domain with the magnetization constrained to point along the islands' long axes.

To construct a candidate ground state for our pattern, first we considered the individual vertices. The energies of all the different configurations of each vertex type in our Penrose tiling in isolation were calculated using the \textsc{oommf} micromagnetic simulation package \cite{oommf}. The lowest and second lowest energy configurations of each type are shown in Fig.~\ref{fig:fig2}a. In each case, the lowest energy magnetic configuration follows a similar `ice rule' to the square pattern and kagom\'{e} pattern, i.e. the macro-spins of each vertex point in and out alternately to minimise the total energy when the co-ordination is even, and take up a configuration as close as possible to this when the co-ordination is odd. In some odd co-ordination vertices, the inability to satisfy this alternating rule, combined with the unequal angles between neighbouring macrospins, gives rise to vertex-level degeneracy. The macrospins where this degeneracy manifests itself are marked with red double-headed arrows in Fig.~\ref{fig:fig2}a. A remarkable feature is that this degeneracy can be suppressed, or emerge, when the vertices are excited into the next energy level.

Having determined the low energy states of the individual vertices, we joined them into the full Penrose tiling using a process of logical steps described in detail in the Supplementary Information. It draws on the idea that the Penrose pattern can be constructed from decagons \cite{1996Gummelt}, although we use a different set of smaller decagons to those in that prior work. First the vertices were joined to form two different types of decagons, which were then fused to form the overall pattern. The result is shown in Fig.~\ref{fig:fig2}b. Two interesting features emerge.

The first is that the full Penrose tiling separates into two parts, which are shown in Fig.~\ref{fig:fig2}b. Whilst some vertex level degeneracies are lifted, new degeneracies arise. Due to the different degree of degeneracy of different types of vertex, some of the macrospins that have twofold degeneracy at the vertex level will lift their degeneracies when connected together and form a rigid framework that spans the entire array. We term this framework the `skeleton', and it is represented by the black spins in Fig.~\ref{fig:fig2}b. It has a unique (up to time-reversal symmetry) ground state, meaning that it possesses long-range order in the sense that specifying the direction of a single macrospin is enough to determine its entire ground state configuration.

The gaps within the skeleton are occupied by clusters of macrospins that are `flippable' as a group with no change in the energy of the whole array in this nearest neighbour picture. These are shown as the groups of red or blue arrows in in Fig.~\ref{fig:fig2}b, and are vertices of types~V, VI, and VII. Those macrospin clusters marked in red surround type~VI or type~VII vertices and have two degenerate time-reversed configurations. Those clusters marked in blue surround type~V vertices, which have a fivefold rotational symmetry. As a result, the macrospin configuration also has a fivefold rotational degeneracy, each configuration of which has a time-reversed degenerate partner. These flippable spins can be expected to be much more susceptible to thermal fluctuations than those in the skeleton (see Supplementary Information). The fact that these clusters do not participate in the long-range ordered skeleton means that the artificial magnetic quasicrystal as a whole possesses macroscopic degeneracy. This picture remains a robust first approximation even with a rigorous treatment of all the long-ranged interactions in the array, since the corrections to the ground state energy are of the order of 0.1\% (see Supplementary Information). Thus our ground state construction procedure explains, through a series of logical steps, the tendency for Monte Carlo simulations of spin systems on quasicrystalline patterns to yield segregated patterns of order and disorder \cite{2003Vedmedenko,2004Vedmedenko,2016Farmer}.

The second is that the global topology of the array means that not every vertex can remain in its vertex-level ground state. Some vertices are forced into their first excited state by `topologically induced emergent frustration', first discovered in the periodic shakti lattice \cite{2014Gilbert}. This arises first at the decagon level, and then more widely when the decagons are joined to form the entire Penrose pattern. In total, 190 out of the 295 Type~\Rmnum{2} vertices in our pattern are forced out of their vertex-level ground state by this mechanism. The locations at which this can happen appear at specific points next to flippable groups, and are marked with the 295 orange circles in Fig.~\ref{fig:fig2}b. These locations have the same five-fold symmetry as the whole pattern, although a specific realisation, such as the example shown by the 190 orange circles that have been filled in Fig.~\ref{fig:fig2}b lower this symmetry. Different frustrated states are realised by flipping all the macrospins in the flippable group, meanwhile the skeleton state remains unique. 

\section*{Macrospin ordering in the skeleton}

\begin{figure*}[t]
  \includegraphics[width=17cm]{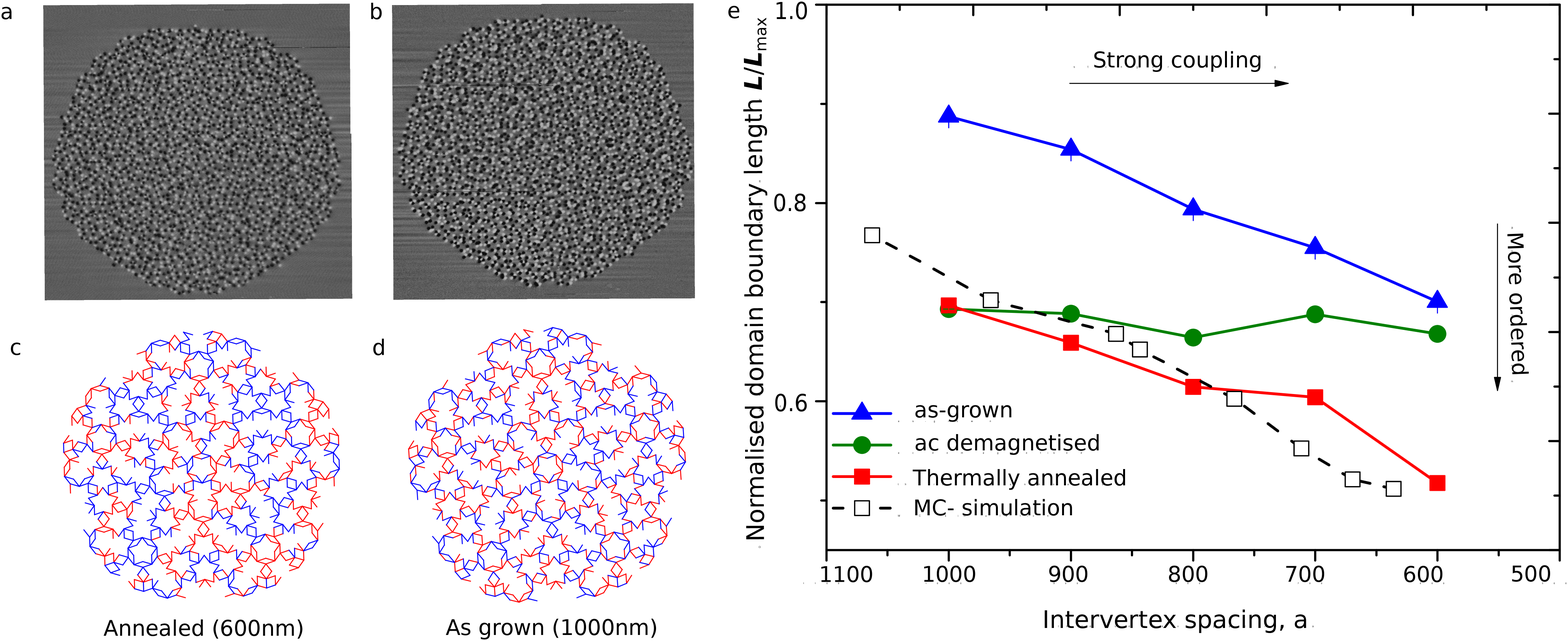}
  \caption{\textsf{\textbf{Ordered regions in the skeleton of an artificial magnetic quasicrystal.} Magnetic force microscopy (MFM) images of an artificial magnetic quasicrystal array \textbf{a}, in a thermally annealed state with $a = 600$~nm and \textbf{b}, in an as-grown state with $a = 1000$~nm. \textbf{c} and \textbf{d}, skeleton macrospin configurations derived from the images in \textbf{a} and \textbf{b}, respectively, compared to the predicted ground state configuration. Red and blue lines correspond to macrospins in the skeleton that match one or the other of the two time-reversed degenerate macrospin configurations in the predicted ground state. Note the longer range of macrospin ordering in the thermally annealed state. \textbf{e}, total number of boundaries between macrospin superdomains $L$ in the skeleton, normalised to $L_\mathrm{max} = 443$, as a function of intervertex distance $a$ (coupling strength) for different state preparation methods. Error bars are smaller than the data markers and so are not shown.}}\label{fig:fig3}
\end{figure*}

The nature of this predicted ground state means that long-range order is expected to manifest itself in the skeleton. We therefore sought experimental evidence for ordering in this part of the array for magnetic configurations that had been prepared by thermal annealing, rotating field AC-demagnetisation, as well as in the as-grown state. We also compared our data with Monte Carlo (MC) simulations in which long-ranged dipolar interactions were considered. Long relaxation times were used to make sure the simulated system was fully equilibrated.

A popular way to identify long-ranged order in artificial spin systems is to use near-neighbour macrospin correlations \cite{2006Wang,2008Ke,2013Zhang}. However, for an aperiodic pattern, in which the immediate environment around each macrospin varies widely from one macrospin to the next, it is not straightforward to define a suitable correlation function. Instead, we used a colour mapping method to identify the macrospin ordering in the skeleton, as shown in Fig.~\ref{fig:fig3}. MFM images of our arrays were acquired for all five values of $a$ following each of three protocols (described in detail in the methods section): the as-grown state, AC-demagnetisation, and thermal annealing.

The skeleton macrospin configurations for the two cases that displayed the greatest ($a = 600$~nm, thermally annealed) and least ($a = 1000$~nm, as-grown) degree of order are shown in Fig.~\ref{fig:fig3}a and Fig.~\ref{fig:fig3}b, respectively. From these, the skeleton macrospin directions were determined and assigned as matching either one or the other of the two time-reversed ground state configurations. These are shown as either red or blue in Fig.~\ref{fig:fig3}c and Fig.~\ref{fig:fig3}d. In the weakly coupled 1000~nm array, the two colours are intermixed in most regions, implying that the system is quite disordered. Whilst ideal long-range order is not found, comparatively large superdomains are found in the strongly coupled 600~nm array in its thermally annealed state. We note that our skeleton is quasi-1D in the sense that it consists of a web of narrow chains of macrospins and so has a reduced local dimensionality. In the 1D Ising model, long-range order can be destroyed by thermal fluctuations at finite temperatures due to the entropy gain outweighing the energy cost. Therefore we expect that the truly long-range order of the ideal ground state may be hard to realise in the skeleton for related reasons.

In order to give a more quantitative description, the total number of quasi-1D domain walls $L$ that separate the macrospin superdomains was determined for each value of intervertex spacing for all three state preparation methods. $L$ is defined as the number of vertices at which elements of different colours meet, and will be smaller if the system is more ordered. Assuming that maximal disorder corresponds to all energy levels having equal occupation numbers, the maximum value of $L$ for the pattern studied here is $L_\mathrm{max} = 443$ (as calculated in the Supplementary Information). The results are illustrated in Fig.~\ref{fig:fig3}e for the three protocols and all values of $a$. As-grown states are the most disordered, especially for more weakly coupled arrays, where $L$ approaches 90 per cent of $L_\mathrm{max}$. Thermally annealed states are the most ordered, especially for the more strongly coupled arrays, although $L$ is never less than about half of $L_\mathrm{max}$. This was not the case with the connected kite-and-dart arrays, in which exchange interactions at the vertices affected the proper thermalisation \cite{Brajuskovic_2016}. AC-demagnetised states show an intermediate level of order, with $L \sim 0.7 L_\mathrm{max}$, that is largely unaffected by the coupling strength. The MC simulations closely match the experimental results for the thermally annealed states, showing that the annealing does indeed provide an excellent thermalisation of the skeleton macrospins. 

\section*{Energy distribution of excited vertices}

\begin{figure*}[t]
    \includegraphics[width=17cm]{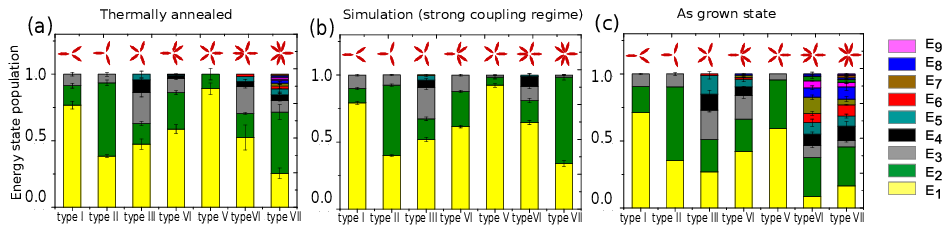}
    \caption{\textsf{\textbf{Populations of the energy levels of the seven vertex types.} Populations for each vertex type for \textbf{a}, the thermally annealed state, \textbf{b}, the MC simulation, and \textbf{c}, the as-grown state are plotted as column stacks. The data in \textbf{a} and \textbf{c} are the average from three patterns with 600~nm intervertex distance. The data in \textbf{b} are averaged from ten configurations obtained by simulation with the strongest interisland coupling that we have tested. The experimental populations for the annealed case strongly resemble the well-thermalised Monte Carlo simulation results.}}
\label{fig:fig4}
\end{figure*}

The presence of the boundaries between superdomains implies that excitations have not been eliminated even in the most highly ordered state that we have prepared. These excited vertices are equivalent to defects in a 1-D Ising model. Based on this, a closer view of the populations of the different skeleton vertex energy levels is of interest, since the energy distribution of excitations above the ground state contains information about the effective thermalisation state of the array \cite{2011Morgan}. These distributions for each type of vertex from the thermally annealed and as-grown states of the $a = 600$~nm array are presented in Fig.~\ref{fig:fig4} and compared with the results of the MC simulations. There is broad agreement between the experimental results for the thermally annealed states and the results of the MC simulations, which we take as further evidence that the annealed states are well-thermalised. The as-grown state, on the other hand, is poorly thermalised. The degree of correlation between the energy level occupancies in the observed and simulated thermalised states was quantified determining the adjusted $R^2$ value, which was only 0.68 for the as-grown state, but as high as 0.97 for the thermally annealed state (see Supplementary Fig.~S\ref{fig:s4_pic1} and accompanying discussion).

A feature that arises for the type~\Rmnum{2} and type~\Rmnum{7} vertices in the thermalised states is that more vertices are found in the first excited state $E_2$ rather than the lowest vertex energy level $E_1$. This occurs not only in the thermally annealed experimental data but also in the simulation. At first sight this appears reasonable for type~\Rmnum{2} vertices, since having every vertex in the lowest energy state cannot be accommodated by the ground state of whole pattern due to topologically induced emergent frustration, as discussed above. (At least 190 out of 295 such vertices in our pattern must stay in $E_2$.)

\begin{figure}[t]
  \includegraphics[width=7cm]{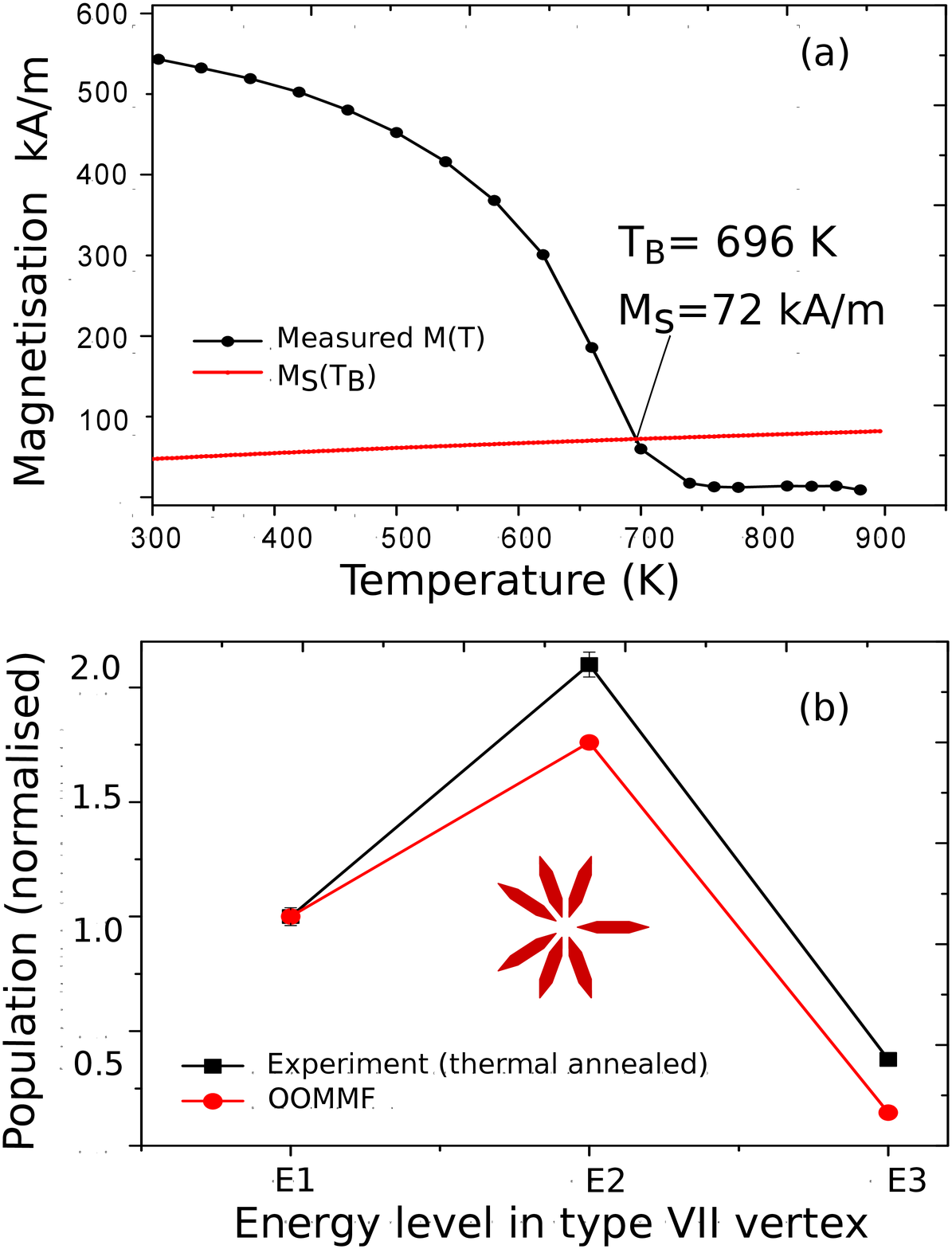}
  \caption{\textsf{\textbf{The blocking temperature and micromagnetic simulation results of the energy population distribution for type VII vertices.} \textbf{a}, the blocking temperature dependence of the calculated saturated magnetisation (see Eq. \ref{eq:blocktemp}) of a nanoisland (red) and the measured magnetisation of a 25~nm thick Permalloy film (black) as a function of  measurement temperature. The intersection of the two curves gives the blocking temperature $T_\mathrm{B}$ and the corresponding saturation magnetisation $M_\mathrm{S}(T_\mathrm{B})$ at that temperature. \textbf{b}, the experimental energy level distribution of type~VII vertices after thermal annealing (black) compared with the micromagnetic simulation results (red). Only the lowest three energy levels occupation are shown. Experimental values are from the MFM images and the simulation results are based on the \textsc{oommf} simulations carried out using parameters relevant at the blocking temperature.}}
  \label{fig:fig5}
\end{figure}

Nevertheless, this argument does not apply to type VII vertices. To understand the causes, we made an estimation of the blocking temperature $T_\mathrm{B}$ for our nanoislands, using the expression \cite{2013Porro,2014Kapaklis}
\begin{equation}
k_\mathrm{B} T_{\mathrm{B}} \ln {\frac{t_\mathrm{m}}{t_0}}=\frac{1}{2} \mu_0 M^2_\mathrm{S}(T_{\mathrm{B}}) D V, \label{eq:blocktemp}
\end{equation}
where $t_\mathrm{m}$ is the measurement time, $t_0$ is the attempt time that is assumed to take a typical value of $10^{-10}$~s,  $\mu_0$ is the permeability of free space, $M_\mathrm{S}$ is the saturation magnetisation, $D$ represents the demagnetising factor of an element, and $V$ is the volume of a single island. In our P3 pattern, where all tile edges have the same length, all islands have nominally identical $V$ and so have nominally the same $T_\mathrm{B}$. This means that they can all thermalise on an equal basis. We took $t_\mathrm{m} = 3$~s (estimated from the cooling rate of thermal annealing experiment of 0.33~K/s), $V = 6.67 \times10^{-22}$~m$^3$ and $D \approx 0.14$, calculated from the expression in Ref. \onlinecite{1945Osborn}. Using this approach, combined with an experimental measurements of the temperature dependence of $M_\mathrm{S}$ for an unpatterned film of Permalloy grown following the same protocol as our artificial magnetic quasicrystal arrays (Fig.~\ref{fig:fig5}a), $T_\mathrm{B}$ was determined to be 696~K, and the saturation magnetization $M_\mathrm{S}(T_\mathrm{B}) = 72$~kA/m at this temperature.

Based on this result, the lowest three configuration energies $E_1$, $E_2$, and $E_3$ of a type~\Rmnum{7} vertex at the blocking temperature were calculated based on the \textsc{oommf} simulation. That calculation showed that the excess population at $E_2$ in both type~\Rmnum{2} and type~\Rmnum{7} can be explained by considering a Boltzmann law $n(j)= g(j) \exp ( -E_j /k_\mathrm{B}T )$, with the contribution at $E_2$ coming mainly from the degeneracy factor $g(j)$ at energy level $j$ (see Fig.~\ref{fig:fig5}b). These degeneracies are given in Fig.~\ref{fig:fig2}a. For both the type~\Rmnum{2} and the type~\Rmnum{7} vertex, the energy difference $E_2-E_1$ is much smaller than for the other vertex types (see Supplementary Information). This feature causes an excess population in the $E_2$ level for modestly elevated temperatures in these two vertex types, and implies that the elimination of excitations is rather difficult as they cool from their annealing. This ease of excitation is enough to explain the excess population on its own in type VII vertices. In type~\Rmnum{2} vertices, topologically induced emergent frustration also forces a certain number of vertices out of their vertex-level ground states, which the small energy spacing easily permits.


\section*{Methods}
\footnotesize{

\textbf{Sample Fabrication.}~Our samples were fabricated from evaporated Permalloy on silicon substrates using a standard electron beam lithography process. The substrates were coated with either thermal oxide, or a 200~nm Si$_3$N$_4$ layer deposited by chemical vapour deposition. The pattern was exposed in ZEP520A resist spun onto the substrate to a thickness of 100~nm, which we then baked at 180$^{\circ}$C for 3 min. We performed the exposure with a JEOL JBX6300-FS writer, with a beam current of 150~pA, a beam step size of 1.75~ nm, and a proximity-effect-corrected (with GenISys Layout BEAMER pattern fracture software) exposure dose. The pattern was developed in amyl acetate at room temperature for 90~seconds, and then rinsed in isopropyl alcohol. We then deposited 25~nm of Permalloy followed by a protective cap of 2~nm of titanium by electron beam evaporation. The pattern was lifted-off in warm n-methyl pyrollidone for several hours, and then rinsed in deionised water and isopropyl alcohol. Sheet films deposited using an identical protocol in the same equipment were characterised using SQUID-vibrating sample magnetometry.

\textbf{Magnetic imaging.}~Magnetic force microscopy (MFM) was the imaging technique that we used in these experiments. Ordering was confirmed using a Veeco Multimode~V, and the main data collection was carried out using a Veeco Nanoman. Both scanning probe microscopes were operated using Bruker MESP tips.

\textbf{As-grown state.}~The samples were first imaged in the state in which they were delivered after fabrication: ``as-grown''. We then subjected them to two demagnetisation procedures in an attempt to produce low-energy states, and imaged them by MFM after each one.

\textbf{AC-demagnetisation.}~The first was an isothermal rotational AC-demagnetisation in a linearly decreasing field, based on the protocols reported in Ref. \onlinecite{2013Morganb}. The sample was securely mounted on a plate attached to a rotating stage and spun at 1,200~r.p.m. The stage was between the pole pieces of an electromagnet, which was used to apply a saturating field of 2,000~Oe that was then ramped to zero at a constant rate of 10~Oe/s, preparing an AC-demagnetised state for imaging.

\textbf{Thermal annealing.}~Our thermal annealing protocol was based on that reported in Ref. \onlinecite{2013Zhang}. All the samples that were subjected to this process were fabricated on the nitride-coated substrates. The annealing process was conducted in an ultrahigh vacuum chamber (pressure less than $10^{-10}$~mbar). In each case, the sample was first heated up to 545~$^{\circ}$C at a rate of 20~$^{\circ}$C per minute, and then held at that temperature for 15 minutes. The sample was then cooled down to 445~$^{\circ}$C at a rate of 1$^{\circ}$C per minute, before temperature control was switched off and the sample allowed to cool naturally. A typical initial rate for this was about 10~$^{\circ}$C per minute. Once the sample had reached room temperature it was retrieved from the vacuum system and the ``thermally annealed'' state was ready for imaging. A square ice sample fabricated under the same conditions as the artificial quasicrystal was fully ordered into its antiferromagnetic ground state \cite{2013Zhang} by this protocol.

\textbf{Micromagnetic simulations.}~The micromagnetic energy hierarchy for vertex configurations at room temperature and the blocking temperature was calculated using the object oriented micromagnetic framework\cite{oommf} (\textsc{oommf}) for islands of the same size and shape as those studied experimentally. The standard Permalloy values of exchange stiffness $A = 1.3 \times 10^{-11}$~J/m, crystalline anisotropy $K = 0$, and magnetisation $M_\mathrm{S} = 8.0 \times 10^{5}$~A/m (at room temperature) and $M_\mathrm{S} = 7.2 \times 10^{4}$~A/m (at the blocking temperature) were used. The damping coefficient $\alpha = 0.5$ was set to an artificially high value to ensure fast convergence of the static micromagnetic states. These energies were used to deduce the candidate ground state using a step-by-step process based on nearest neighbour interactions.

\textbf{Monte Carlo simulations.}~We performed Monte Carlo simulations of a 1650-island Penrose ice pattern. Each island was represented by an Ising spin positioned at its centre and with anisotropy direction parallel to the island long axis. Each spin interacts with all other spins in the system via dipolar interactions, that is, $U_{ij} = d (\mathbf{\hat{m}}_i \cdot \mathbf{\hat{m}}_j - 3(\mathbf{\hat{m}}_i \cdot {\mathbf{\hat{r}}}_{ij})(\mathbf{\hat{m}}_j \cdot \mathbf{\hat{r}}_{ij}))/|\mathbf{r}_{ij}|^3$ with $d=\mu_0 m^2/8\pi$ controlling the strength of interactions, where $\mathbf{\hat{m}}$ is a unit vector in the direction of the magnetic moment $m$ of each island, and $\mathbf{r}_{ij}$ is the displacement vector between the dipole moments $\mathbf{\hat{m}}_i$ and $\mathbf{\hat{m}}_j$. In all simulations reported in the main text, the system was initialized in a reproducible microstate, with all spins having a positive projection on the $x$-axis. It was then evolved using a standard Monte Carlo algorithm where each single spin flip is accepted with probability $\exp(-\beta \Delta U)$, until a steady state is obtained. For an inter-island spacing of 600 nm, simulations were performed with $\beta e_0=0.814$, where $e_0$ is an energy scale defined by the energy difference between levels 1 and level 2. The \textsc{oommf} simulations gave a value of $9\times10^{-21}$~J for $e_0$ for the 600~nm spaced system, so that in these simulations $T \approx 800$ K, comparable in magnitude to the annealling temperature that we used. An inter-island spacing of 1050 nm corresponds to $\beta e_0=0.174$. These simulations were used to show that the nearest neighbour picture is a very close approximation to one in which all interactions are considered.
}

\section*{References}
\bibliographystyle{naturemag}

\bibliography{ASI_refs}

\vskip 1cm
\section*{Acknowledgements}
This research used resources of the Center for Functional Nanomaterials, which is a U.S. DOE Office of Science Facility, at Brookhaven National Laboratory under Contract No. DE-SC0012704, and was also supported by the EPSRC (grant EP/L00285X/1). ZB was supported by the ERC Advanced Grant no. 291002 SIZEFFECTS.

\section*{Author Contributions}
C.H.M. conceived the project. A.S. fabricated the samples, using Penrose array designs made by G.B. and D.S. D.S. performed the MFM imaging and the AC-demagnetisation and thermal annealling protocols, analysed the data (using software co-designed with G.B.), and theoretically derived the expected ground state. Z.B. performed the Monte Carlo simulations in consultation with D.S. S.A.M. performed the SQUID magnetometry measurements. C.H.M., G.B., and P.D.O. supervised the work. All authors discussed the data and commented on the manuscript.

\section*{Data Availability Statement}
The datasets generated during and/or analysed during this study are publicly available in the University of Leeds Research Data Repository, https://doi.org/10.5518/176.

\section*{Additional Information}
Supplementary information is available in the online version of the paper. Reprints and permissions information is available online at www.nature.com/reprints.

Correspondence and requests for materials should be addressed to C.H.M.

\section*{Competing financial interests}
The authors declare no competing financial interests.

\end{document}